\documentclass[10pt,conference,letterpaper,twocolumn]{IEEEtran}

\IEEEoverridecommandlockouts
\usepackage{sansmathaccent}
\pdfmapfile{+sansmathaccent.map}
\usepackage[T1]{fontenc} 
\usepackage{amsmath,amsfonts}
\usepackage[cmintegrals]{newtxmath}
\usepackage{bm} 
\usepackage[noadjust]{cite}
\usepackage{caption}
\usepackage{xcolor}
\usepackage[utf8]{inputenc}
\usepackage{ragged2e}
\usepackage{graphicx}
\graphicspath{ {./images/} }
\usepackage{mathtools}

\usepackage{calrsfs}
\usepackage{textcomp}
\usepackage{epstopdf}
\usepackage{epsfig}
\usepackage{xcolor}
\ifCLASSOPTIONcompsoc
\usepackage[caption=false,font=normalsize,labelfont=sf,textfont=sf]{subfig}
\else
\usepackage[caption=false,font=footnotesize]{subfig}

\DeclareMathAlphabet{\pazocal}{OMS}{zplm}{m}{n}

\newcommand{\Fb}{\pazocal{F}}
\newcommand{\Nb}{\pazocal{N}}

\newcommand{\Gb}{\pazocal{G}}
\newcommand{\Lb}{\pazocal{L}}
\newcommand{\Sb}{\pazocal{S}}

\usepackage[linesnumbered, lined, ruled]{algorithm2e}
\SetKwInOut{Parameters}{Parameters}
\let\oldnl\nl
\newcommand{\nonl}{\renewcommand{\nl}{\let\nl\oldnl}}
\makeatletter
\newcommand{\removelatexerror}{\let\@latex@error\@gobble}
\makeatother

\usepackage[pdfencoding=auto, psdextra]{hyperref}

\begin{document}	

\title{Low Latency Scheduling Algorithms for Full-Duplex V2X Networks\\}

\author{\IEEEauthorblockN{Michail Palaiologos\textsuperscript{*\dag}, Jian Luo\textsuperscript{*}, Richard A. Stirling-Gallacher\textsuperscript{*} and Giuseppe Caire\textsuperscript{\dag}}
	\IEEEauthorblockA{\textsuperscript{*}Munich Research Center, Huawei Technologies Duesseldorf GmbH, 80992 Munich, Germany\\ 
	Email:\{michail.palaiologos, jianluo, richard.sg\}@huawei.com}
\IEEEauthorblockA{\textsuperscript{\dag}Communications and Information Theory Group, Technische Universit{\"a}t Berlin, 10587 Berlin, Germany\\
Email:{caire@tu-berlin.de}}
}

\maketitle

\begin{abstract}
Vehicular communication systems have been an active subject of research for many years and are important technologies in the 5G and the post-5G era. One important use case is platooning  which is seemingly the first step towards fully autonomous driving systems. Furthermore, a key performance parameter in all vehicular communication systems is the end-to-end packet latency. Towards this goal, full-duplex (FD) transceivers can potentially be an efficient and practical solution towards reducing the delay in platooning networks. In this paper, we study the delay performance of dynamic and TDMA-based scheduling algorithms and assess the effect of FD-enabled vehicles with imperfect self-interference cancellation (SIC). By simulations, we demonstrate the performance-complexity trade-off of these algorithms and show that a TDMA centralized scheme with low-complexity and overhead can achieve comparable performance with a fully-dynamic, centralized algorithm.
\end{abstract}


\section{Introduction}\label{SecI}
\begin{figure}[!t]
	\centering
	\includegraphics[width=\columnwidth, height = 1.5cm]{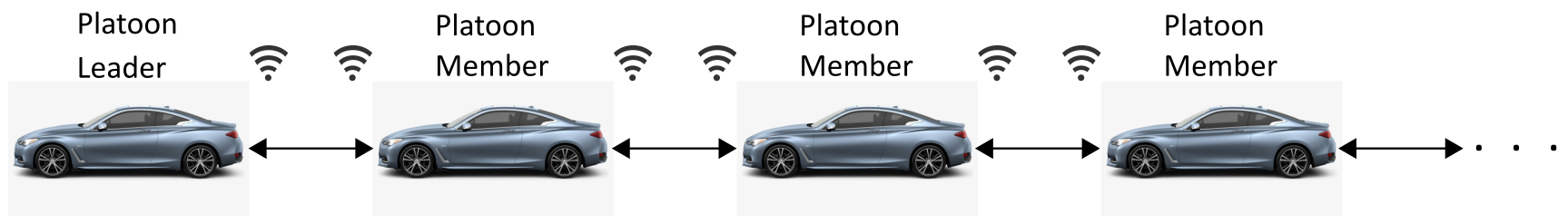}
	\caption{Vehicular platoon network}
	\label{fig:platoon}
\end{figure}

Vehicular communications, most commonly refered to as vehicle-to-everything (V2X) communication systems, is one of the most promising applications of fifth-generation (5G) systems that have the potential to dramatically increase the efficiency and safety of transportation. By exchanging valuable information, vehicles can interact with each other in a way that the probability of collision can be reduced compared to todays transportation systems. One of the main applications of V2X communications is platooning. A platoon is a group of vehicles where each vehicle is located behind another, so that they all together form a chain of vehicles (Fig. \ref{fig:platoon}).

The exchange of control messages wirelessly between neighboring vehicles has to be at a high rate and with ultra low latency, so that the danger of platoon instability is minimized \cite{jia2015survey}. In order to achieve low end-to-end packet latency\footnote{End-to-end latency is the time that a packet spends in the network, starting from the time of arrival until the time of departure from the network} values, FD-enabled vehicles could be introduced as a potential solution. In particular, vehicles having distributed transceivers at different parts of the car body can easily realize full-duplex transmission, i.e. transmit and receive simultaneously at the same frequency.

It has already been demonstrated that FD can be leveraged in V2X scenarios towards increasing the sum-rate throughput of the network as well as reducing the latency \cite{7946944}. However, the majority of the related literature is focused either on the outage probability or the throughput performance of these networks by distributed resource allocation. In this paper, a centralized resource allocation problem for a platoon network under the coverage of a base-station (BS) is studied. We focus solely on the latency performance associated with the communication between the platoon vehicles and not between the BS and the platoon leader (PL). Furthermore, it is assumed that the serving BS assigns orthogonal resources to neighbouring platoons.

There have been many works in the past that have studied centralized scheduling policies and proposed different solutions \cite{1498534, 182479}. Recently, there have also been studies where different scheduling algorithms are implemented for relay networks, where line networks that are similar to platoon networks are examined \cite{7946944, song2020queue}. However, the assumption in most of the existing literature is that the network nodes do not have full-duplex (FD) capabilities.

In this work, we study the effect of FD-enabled vehicles in the latency performance of a platoon network under the presence of imperfect self-interference cancellation (SIC). Furthermore, we propose two low-complexity and low-overhead TDMA-based scheduling schemes that allocate resources per frame (frame-based scheduling) and compare their performance with a slot-based scheduling algorithm as a benchmark.

The remainder of the paper is organized as follows. Section II describes the system model. In Section III different scheduling algorithms are presented where the effect of FD is also taken into account. Section IV deals with the performance evaluation of the scheduling algorithms and, finally, Section V contains concluding remarks and suggestions for future work.

\section{System Model}\label{SecII}
\begin{figure}[!t]
	\centering
	\includegraphics[width=\columnwidth, height = 1.5cm]{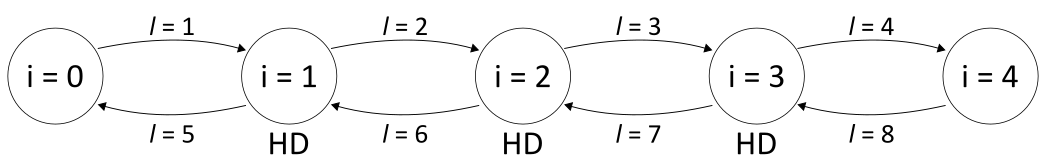}
	\caption{Platoon network graph model}
	\label{fig:graph}
\end{figure}

Fig. \ref{fig:graph} shows a directed graph representation $\Gb = \{\Nb, \Lb\}$, where $\Nb$ and $\Lb$ represent the set of vertices and edges of the graph, respectively. Let $N = |\Nb|$ and $L = |\Lb|$ be the number of vertices (nodes) and edges (links) in the network respectively (we use $|\cdot|$ to represent the cardinality of a set), while variables $i$ and $l$ denote the elements of sets $\Nb$ and $\Lb$. $\Nb^{HD}$ and $\Nb^{FD}$ represent the set of half-duplex (HD) and FD vehicles, respectively. We allow arbitrary number and positions of HD and FD nodes. $N_r$ represents the number of platoon members (PMs), i.e. $N_r = N - 2$. The terms node and vehicle are used interchangeably in this paper.

Typically, a PM has to exchange information with the PL and with its preceding vehicle. Towards this, the underlying assumption is that several flows are present in this network, where each flow is associated with a distinct transmitter - receiver pair, as illustrated in Fig. \ref{fig:allflows}. The existence of flows between the PL and the platoon tail vehicle in this flow model allows us to evaluate the largest possible end-to-end latency on a platoon network. Let $\Fb$ represent the set of existing flows and $S(f)$ the source node of flow $f$, where $f \in [1:F]$ and $|\Fb| = F$. Half of these flows are related to the transmission of packets from the PL to all PMs. Due to the distinct direction of the information flow of these flows we will refer to them as the "right-hand" flows. The other half of the flows will be refered to as the "left-hand" flows. The assumption is that a vehicle can communicate directly only with its \textit{neighbouring} vehicles, i.e. the ones that are only one-hop away (Fig. \ref{fig:graph}) due to the low transmission power and the blocking effect of the vehicles at the selected frequency.

It is assumed throughout that time is slotted. The capacity of each link $l$ at time slot $t$ is denoted as $C_l(t)$. The transmission rate of flow $f$ over link $l$ at time slot $t$ is given as $\mu_l^f(t)$. At each time slot, packets from only one flow can be transmitted through an activated link. Due to the underlying flow model, routing constraints should be introduced, so that not all flows are allowed to use all links. If the set of links that flow $f$ is allowed to use is denoted as $\Lb_f$, then for all "right-hand" flows, i.e. flows $f \in [1:\frac{F}{2}]$ we have that $\Lb_f = \{l: l \leq f\}$. For the "left-hand" flows, i.e. $f \in [\frac{F}{2}+1 : F]$ the set of links they are allowed to use is given by $\Lb_f = \{l: \frac{L}{2} + 1 \leq l \leq f\}$. Hence, $\mu_l^f(t) = 0 \: \forall \, t$, if $l \notin \Lb_f$. 
\section{Scheduling Schemes}\label{SecIII}
\subsection{TDMA Scheduling - Flow-based (FB)}
\begin{figure}[!t]
	\centering
	\includegraphics[width=\columnwidth, height = 2.4cm]{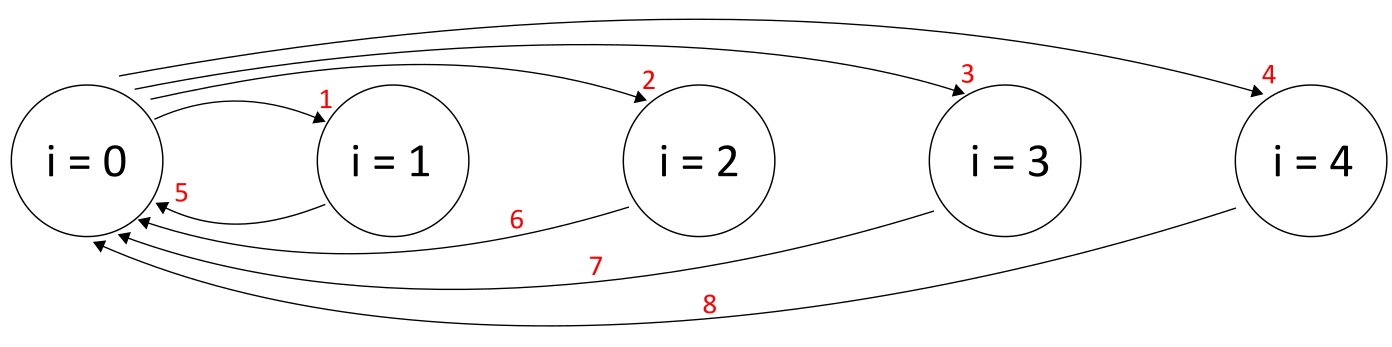}
	\caption{Flow model - Numbers in red indicate the flow index}
	\label{fig:allflows}
\end{figure}

Usually the goal in most of the publications related to TDMA scheduling is to obtain the minimum possible number of slots (per frame), as in \cite{1498534}. A TDMA scheduling approach with the minimum frame length for the underlying platoon network activates one link per frame under the assumption of \textit{one-hop interference}. However, as it is illustrated in Fig. \ref{fig:allflows}, links have to support a varying number of flows, depending on their position in the network. Therefore there is a varying level of (flow) "congestion" at each link. In order to model this effect, let $\Sb_t$ represent the set of links that are co-scheduled for transmission in slot $t$ and $T$ represent the frame length in number of slots. Then, the number of slots that each link $l$ is scheduled per frame is given as
\begin{equation}\label{eq:5}
o_l = \sum_{t = 1}^{T} I\{l \in \Sb_t\}
\end{equation}
where $I\{\cdot \}$ is the indicator function (which returns 1 if the argument inside its brackets is true, otherwise it returns 0). In order to model the congestion in each link, we use variable $d_l$ which is equal to the number of flows that have link $l$ in their routing path and is given as
\begin{equation}\label{eq:6}
d_l = \sum_{f = 1}^{F} I\{l \in \Lb_f\}
\end{equation}

In order to take the traffic pattern into account, the number of slots per link $o_l$ needs to be proportional to the congestion of this link $d_l$. In the case of a platoon network, where the number of flows $F$ is small, the number of slots per link can be exactly equal to the congestion level of this link, so that $o_l = d_l$. In this way, the slot allocation of each directional link $l \in \Lb$ can be derived very simply from Fig. \ref{fig:allflows}. Particularly, an edge coloring technique similar to the one in \cite{1498534} can be used towards assigning a specific number of colors (slots) to each link. In the case of our network, the topology in Fig. \ref{fig:graph} is a bipartite graph. It was proven in \cite{konig1916graphen} that the minimum number of colors required to color all edges of a bipartite graph is equal to $\Delta$, where $\Delta$ is the maximum degree of the graph. Thus, a per-link color allocation can be easily implemented at the beginning of each frame, where the number of flows that are incident to each node dictate the degree of this node and, hence, the exact edge color allocation. 

\begin{figure}[!t]
	\centering
	\includegraphics[width=\columnwidth, height = 1.4cm]{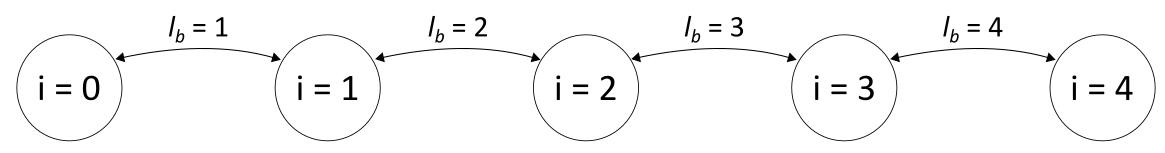}
	\caption{Bidirectional links}
	\label{fig:bidiLinks}
\end{figure}

Instead of assigning a set of colors to each directional link $l$, we could assign the same set of colors to each \textit{pair} of directional links that connect two neighbouring nodes. Consequently, the color assignment specifies only the bidirectional link that needs to be activated in each slot. In Fig \ref{fig:bidiLinks}, $l_b$ is the index of bidirectional links of the platoon network and let $\Lb_b$ represent the set that contains all these links.
This modification allows an extra degree of freedom on the \textit{flow scheduling}, since the flow that is activated in every slot can be decided after the two nodes that are the endpoints of the activated bidirectional link exchange information on the sizes of their queue backlogs.

\begin{figure}[!t]
	\centering
	\includegraphics[width=\columnwidth, height = 1.5cm]{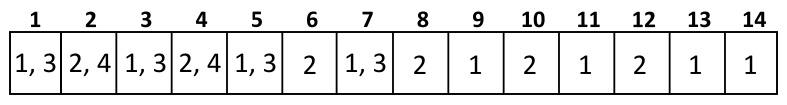}
	\caption{Frame structure - Bidirectional links activation}
	\label{fig:frameStr}
\end{figure}

Given the bidirectional-links network model in Fig. \ref{fig:bidiLinks} and the traffic pattern that is illustrated in Fig. \ref{fig:allflows}, the number of colors per bidirectional link is given in a vector format as
\begin{equation}\label{eq:9}
\underline{o}_{l_{b}} = 
\begin{bmatrix}
o_{1} & o_{2} & o_{3} & o_{4}
\end{bmatrix}
= 
\begin{bmatrix}
8 & 6 & 4 & 2
\end{bmatrix}
\end{equation}
where the $l_b$ - th element of this vector is the number of colors allocated to bidirectional link $l_b$ and is given by
\begin{equation}\label{eq:10}
o_{l_{b}} = 2(N_r + 1) - 2(l_b - 1)
\end{equation}
Therefore, due to \cite{konig1916graphen} and under the one-hop interference model, the total number of colors (slots) required to support this color allocation is equal to
\begin{equation}\label{eq:11}
\Delta = \max_{1 \leq l_{b} \leq |\Lb_b|-1} \{o_{l_b} + o_{l_{b} + 1}\} = 8 + 6 = 14
\end{equation}
So, according to this slot allocation, the frame consists of $T = 14$ slots and is demonstrated in Fig. \ref{fig:frameStr}, where the bidirectional links activated in each slot are also demonstrated.

\subsection{Back-pressure}
In the previous algorithm, although link congestion is taken into account, there is no consideration for the backlog sizes of the queues. In a V2V platoon environment, where vehicles are expected to transmit in bursts due to random changes of the surrounding environment, an efficient scheduling algorithm ought to take into account the size of the backlogs. In order to incorporate the queue backlog information in the scheduling process, we implemented the back-pressure (BP) algorithm \cite{182479}. BP can be modeled as a binary integer programming problem, where a binary vector is introduced to capture the link activations on a slot basis. In particular, this binary scheduler vector $\pmb{\Lambda}(t) \in \mathbb{R}^{L}$ is updated every slot as the solution of the following optimization problem
\begin{equation}\label{eq:12}
\pmb{\Lambda}(t) = \arg \max_{\Lambda_{l}} \sum_{l=1}^{2(N+1)} W_{l}(t) \cdot \mu_{l}(t) \cdot \Lambda_{l}(t)
\end{equation}
Link $l$ is activated in slot $t$ only if the solution of the optimization problem returns $\Lambda_l(t) = 1$. $W_{l}$ represents the queue differential backlog. 

As flows have a different number of hops in their path, their latency performance will not be the same when their input rates are equal. In order to deal with these fairness issues, the computation of the queue differential backlogs $Q$ of the flows may depend on their hop count. To this extent, regarding the queue backlog of each flow, instead of being raised to the power of 1 for all flows, this exponent can vary depending on the number of hops in each flow \cite{5688207}. If we use $\gamma_{f}$ to represent the exponent of flow $f$, then
\begin{subequations}
	\begin{align} 
	W_l(t) = \max_{f} \{(Q_{i - 1}^f)^{\gamma_{f}} - (Q_i^f)^{\gamma_{f}}, 0\} \label{sub-eq-14:1}\\
	W_l(t) = \max_{f} \{(Q_{i}^f)^{\gamma_{f}} - (Q_{i-1}^f)^{\gamma_{f}}, 0 \} \label{sub-eq-14:2}
	\end{align}
\end{subequations}
where $i \in [1 : N + 1]$ and $l \in [1 : \frac{L}{2}], \, f \in [1 : \frac{F}{2}]$ for "right-hand" flows (\eqref{sub-eq-14:1}) and $l \in [\frac{L}{2} + 1 : L], \, f \in [\frac{F}{2} + 1 : F]$ for "left-hand" flows (\eqref{sub-eq-14:2}). In these two expressions, link $l$ is incident to nodes $i$ and $i-1$. Finally, the destination node of a flow does not keep an internal queue dedicated to this flow.

In order to resolve the conflicts of the network, a few constraint inequalities need to be added to the optimization problem of \eqref{eq:12}. These constraints, which can be expressed via the binary scheduler vector $\Lambda$, basically capture the edge coloring procedure on the topology graph.
\begin{subequations}
	\begin{align} 
	&\mu_l(t) = \text{min} \{Q_{i}^f(t), C_l(t)\}, \, \forall \, i \in \Nb, \: \forall \, f \in [1:F] \label{sub-eq-15:1}\\
	&\Lambda_{l}(t) \in \{0, 1\}, \, \forall \, l \in \Lb \label{sub-eq-15:2}\\
	&\sum_{l \in \{i, i+1, i+N_r+1, i+N_r+2\}} \Lambda_{l}(t) \leq 1, \, \forall \, i \in \Nb^{HD} \cap [1:N_r], \notag \\
	&\forall \, l \in \Lb \label{sub-eq-15:3}\\
	&\sum_{l \in \{1, N_r+2\}} \Lambda_{l}(t) \leq 1, \, \text{if} \; i = 0 \in \Nb^{HD} \label{sub-eq-15:4}\\
	&\sum_{l \in \{N_r+1, 2N_r+1\}} \Lambda_{l}(t) \leq 1, \, \text{if} \; i = (N_r+1) \in \Nb^{HD} \label{sub-eq-15:5}\\
	&\sum_{l \in \{i+1, i+N_r+1\}} \Lambda_{l}(t) \leq 1, \, \forall \, i \in [1:N] \label{sub-eq-15:6}\\
	&\sum_{l \in \{i, i+N_r+2\}} \Lambda_{l}(t) \leq 1, \, \forall \, i \in [1:N] \label{sub-eq-15:7}\\
	&\mu_l^f(t) = 0 \: \forall \, t, \, \text{if} \; l \notin \Lb_f \label{sub-eq-15:8}
	\end{align}
\end{subequations}
In particular, constraints \eqref{sub-eq-15:3} - \eqref{sub-eq-15:5} are related to the HD constraint, while constraints \eqref{sub-eq-15:6} and \eqref{sub-eq-15:7} are necessary in order to assure that no node is allowed to transmit (receive) to (from) more than one other nodes at any given time slot. Finally, \eqref{sub-eq-15:1}, \eqref{sub-eq-15:2} and \eqref{sub-eq-15:8} are due to the system model.
 
\subsection{TDMA Scheduling - Queue-based (QB)}
In practise, it is rather difficult to perform an adaptive algorithm such as BP. Additionally, the signalling overhead of an adaptive algorithm such as BP is very large, as information need to be exchanged between the vehicles and the BS at every slot. Note that the main advantage of BP is that it takes into consideration variations in queue backlogs and channel conditions at every slot. But, in a platoon environment, where all links are LOS, the variations of the link capacities are usually small from one slot to the next. Thus, one reasonable assumption could be that link capacities remain the same during each frame. In order to combine the advantage of the low overhead of TDMA schemes and the dynamic behavior of an algorithm such as BP, we introduce a TDMA scheduling scheme that utilizes the information of queue backlogs at the beginning of each frame before the slot allocation.

More formally, the link that maximizes an objective function (similar with BP) can be obtained for each flow $f$ as
\begin{equation}\label{eq:16}
\displaystyle\arg \max_{l} \{\frac{W_l^f(t_S)}{\mu_l(t_S)} \}
\end{equation}
where $t_S$ is the begining of the $S$-th time frame. This approach indicates that at least one slot is allocated per flow $f$ for the upcoming frame, given that there is at least one non-empty queue in this flow's path. Then, over these $F$ values, the number of times that each link $l$ is the solution of this optimization problem is the "demand" of each link, which is actually the number of colors that need to be assigned to this link. In order to calculate the demand of each bidirectional link $l_b \in \Lb_b$, simply add the demand of each directional link $l_b$ with the demand of its "mirror" link $l_b + N_r + 1$. As a flow can have at maximum $L/2$ links in its path (Fig. \ref{fig:allflows}), the computational complexity of this step, which has to run at the beginning of each frame only, is of the order of $O(F \cdot L)$. So, each node has a specific degree and the edge coloring can be carried out based on the demand of the bidirectional links. 

If the size of the frame is equal to $T$ slots, then the actual number of allocated slots is equal to $T_{act} = \text{min}\{T, \Delta\}$, where $\Delta$ is the maximum degree of the graph, after the demand of each link has been determined. If $\Delta > T$, the demand of the bidirectional links needs to be reduced (starting from the links with the highest demand) until $\Delta = T$. If $\Delta < T$, the demand of the bidirectional links needs to be increased in, e.g., a sequential manner, until $\Delta = T$. Finally, in \eqref{eq:16} we also divide the differential backlogs with the link capacity of each link at the beginning of each frame. Since the assumption is that link capacities remain the same during each frame, this action affects slightly the results. The rationale behind it is that we assign extra priority to links with bad channel conditions.

\subsection{Full-Duplex Consideration}
The case of heterogeneous platoon networks, where both HD and FD nodes exist, requires special attention, since the conflict-free independent sets that graph's edges belong to are different when the FD nodes positions vary. For example, if node $i = 1$ is FD, then bidirectional links $l_b = 1$ and $l_b = 2$ can belong to the same independent set and basically form a single "FD link". As a result, these two links can be treated as one and can be assigned the same colors. So, if the number and/or the position of FD nodes in the network changes, the color assignment of the bidirectional links is also changed. 

Note that the TDMA schemes return results for the activation of bidirectional links only and do not include any information for the scheduling of flows. At each slot, for each activated bidirectional link, one of the contending flows over this link can be scheduled based on its queue differential backlog. If an "FD" bidirectional link is activated, there are at least three nodes involved in the scheduling decision, so there are more queues involved. In order to satisfy this, the first step would be to calculate the max differential backlog over all directional links. Then, instead of searching and comparing between the max differential backlog of the two "types" of flows (as in the HD case), first we need to sum the max differential backlogs of all flows that belong to the same "type" and then compare these added differential backlogs in order to determine if a "right-hand" or "left-hand" flow will be scheduled. Hence, a more fair flow scheduling is achieved.

\section{Numerical Results}\label{SecIV}
In this section, the performance of the previously described scheduling algorithms for a platoon network with both HD and FD vehicles is evaluated. The parameters that were used for the simulations are outlined in Table \ref{table1}. An example size platoon of 5 vehicles is evaluated and a line-of-sight channel model with associated path loss is assumed between adjacent platoon nodes. The frame length is fixed for fixed HD and FD locations and is given by Fig. \ref{fig:frameStr}.
\begin{table}[!t]
	\renewcommand{\arraystretch}{1.2}
	\caption{Simulation Parameters}
	\label{table1}
	\centering
	\begin{tabular}{|c|c|}
		\hline
		\bfseries Parameter & \bfseries Value \\
		\hline
		Platoon size, N & 5 \\
		\hline
		Number of flows, F & 8 \\
		\hline
		Vehicle length and separation & 5 m and 33.33 m \\
		\hline
		Carrier frequency and bandwidth & 30 GHz and 200 MHz \\
		\hline
		Transmit power & 23 dBm \\
		\hline
		Shadowing: mean, standard deviation & 0 mean, 8 dB standard deviation \\
		\hline
		Number of slots and slot duration & 40000 slots with 125 $\mu s$/slot \\
		\hline
		Self-interference cancellation level & 10 dB, 40 dB \\
		\hline
		Rate of Poisson arrivals, $\lambda$ & 0.04 packets/slot \\
		\hline
	\end{tabular}
\end{table}
\begin{figure}[!t]
	\centering
	\subfloat[]{\includegraphics[width=8.5cm, height=4.5cm]{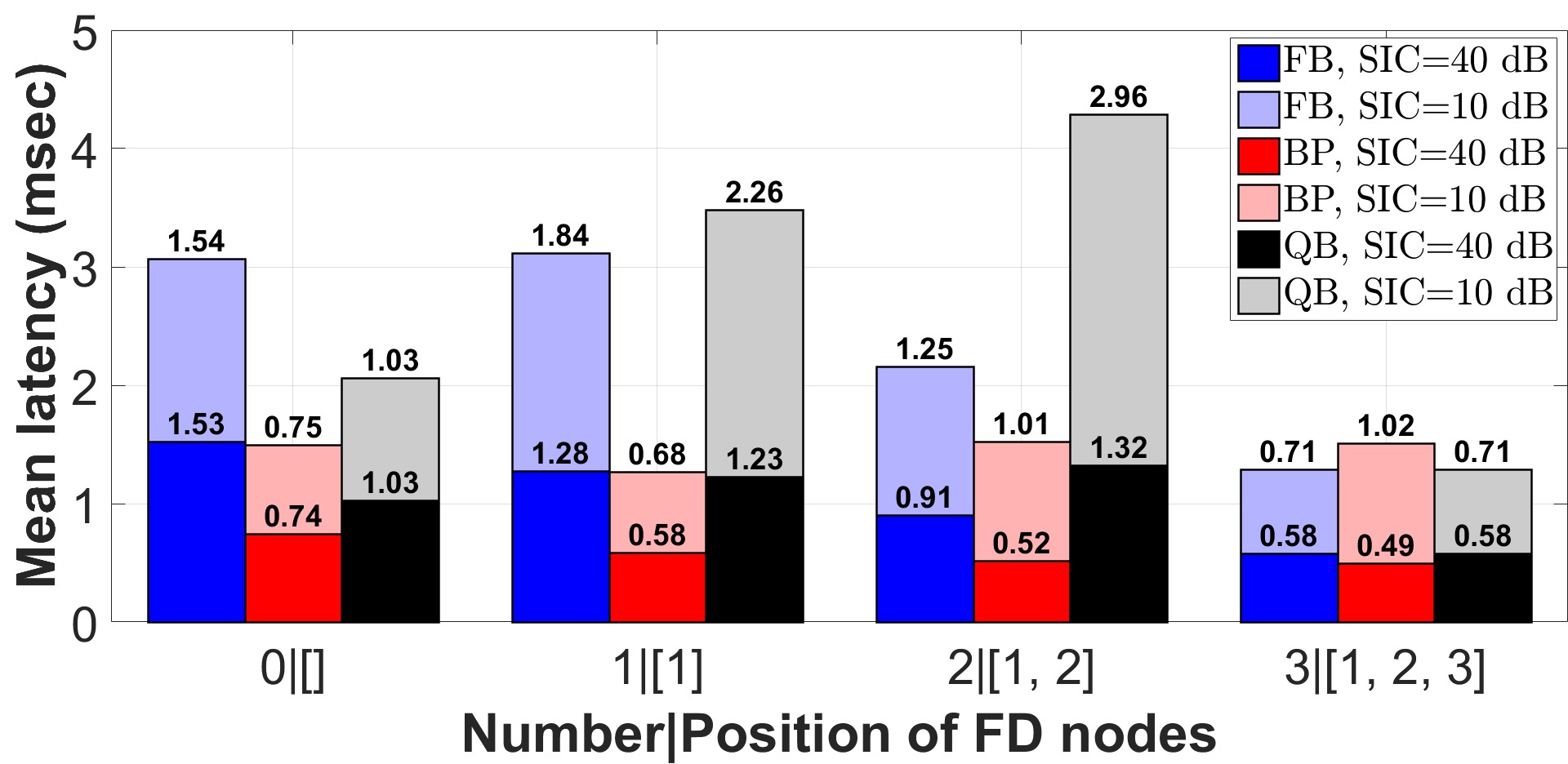}
		\label{meanLat_0.04}}
	\hfil
	\subfloat[]{\includegraphics[width=8.5cm, height = 4.5cm]{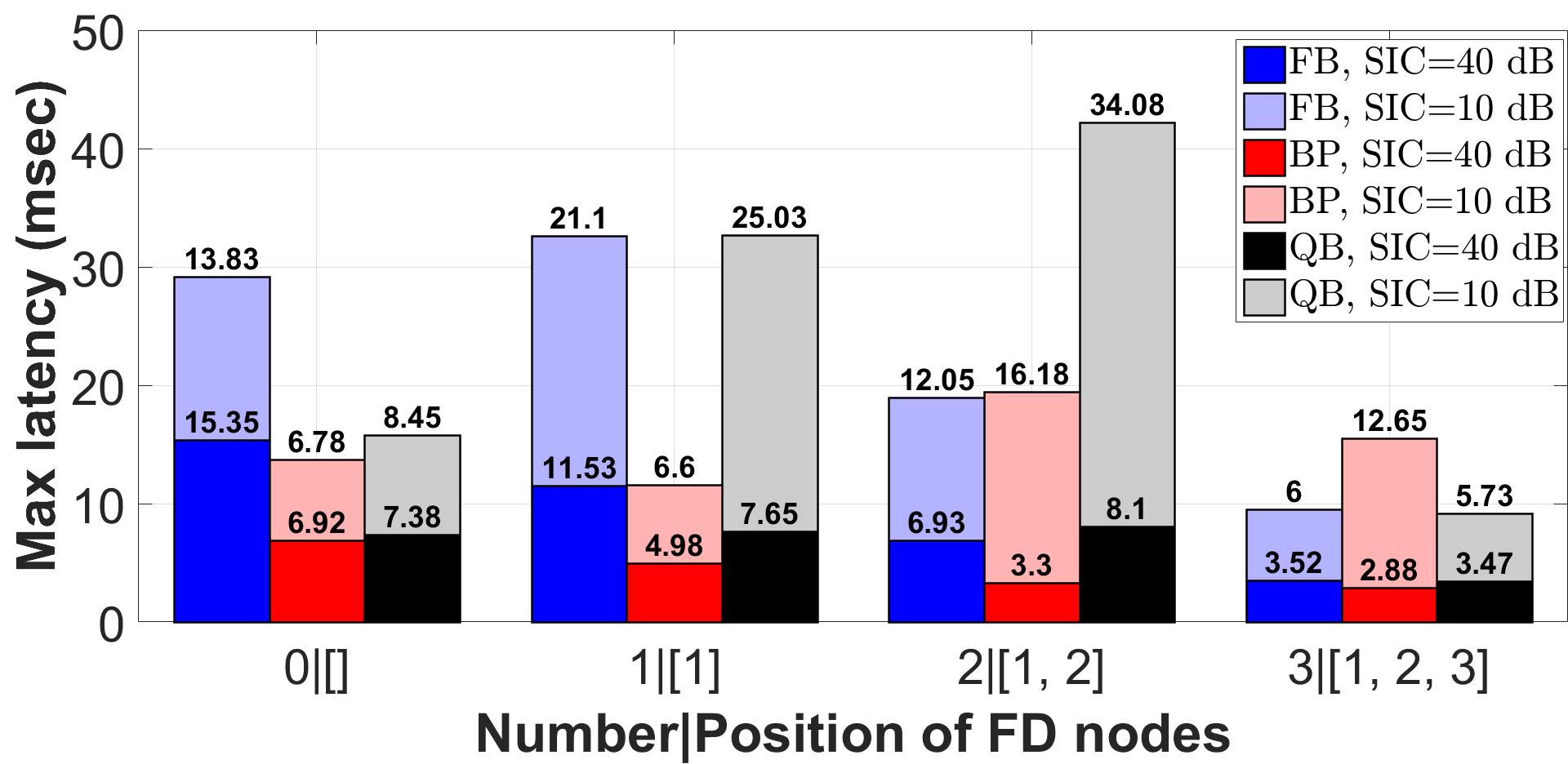}
		\label{maxLat_0.04}}
	\caption{(a) Mean and (b) max latency for 2 different levels of SIC and an input rate of 0.04 packets/slot}
	\label{latency}
\end{figure}
In Fig. \ref{latency}, the mean and max end-to-end packet latency is demonstrated under the three scheduling algorithms, for varying number and position of FD nodes when the arrival rate is equal to $0.04$ packets/slot. The actual values of the input rates are derived from the flow control process in \cite{doi:10.2200/S00271ED1V01Y201006CNT007}, so that network stability is guaranteed. The chosen arrival rates for all flows are assumed to be equal and strictly inside the capacity region. The arriving packet sizes are uniformly distributed in the range of $[40:32:136]$ kbits. Two distinct SIC levels are also studied.

In Fig. \ref{latency}, as expected, in the case where the SIC is at an acceptable level (i.e. equal to 40 dB) an algorithm such as BP that takes scheduling decisions at every slot performs much better compared to the two TDMA scheduling schemes. However, the average delay of all these schemes remains low for a SIC of 40 dB, as it is not more than 1.53 msec. Possibly more valuable for practical purposes is the information related to maximum latency in Fig. \ref{maxLat_0.04}. It can be seen that a TDMA sheduling scheme that takes queue backlogs into account (QB scheme) delivers latency performance close to the BP scheme and significantly better than the FB algorithm when there are no FD vehicles or a single FD vehicle, while achieving that with much lower complexity and signalling overhead compared to BP. 

Note that although the QB algorithm outperforms the FB algorithm, when a single FD vehicle is in position 1, its latency performance becomes worse when adding more FD vehicles. The reason is that this algorithm does not take into account the varying congestion levels in the links and therefore over-schedules links with low congestion levels. In contrast, a simple scheme such as the FB algorithm, that takes into account the congestion can only really take advantage of the existence of FD nodes.

It can also be seen from these two figures that, as shown in \cite{7946944}, relatively low levels of SIC are sufficient for satisfactory latency performance in V2V networks, since, in practise, SIC of around 40 dB could be a reasonable assumption. Nonetheless, there is a limit on the acceptable levels of SIC. In fact, we can see that, even for a dynamic algorithm like BP, its performance deteriorates for SIC = 10 dB when FD vehicles are added. On the other hand, FB scheme is very robust, since its latency performance is improved when adding FD vehicles, even when the SIC is equal to 10 dB.

Regarding the weights $\gamma_{f}$ that have been used in the differential backlogs expressions in \eqref{sub-eq-14:1} and \eqref{sub-eq-14:2}, they vary depending on the number of hops in each flow. It is not easy to come up with a deterministic solution for these weights due to the increased coupling between the flows \cite{5688207}. Given the hop-count, a reasonable choice for flows with one hop in their path would be $\gamma_{f} = 0.8$, while for flows with 2 and 3 hops, a value of $\gamma_{f} = 0.9$ appears to produce better results. Finally, for flows with the max possible number of hops, the exponent used in their differential backlog expressions is equal to 1.
\begin{figure}[!t]
	\centering
	\includegraphics[width=8.5cm, height = 4.5cm]{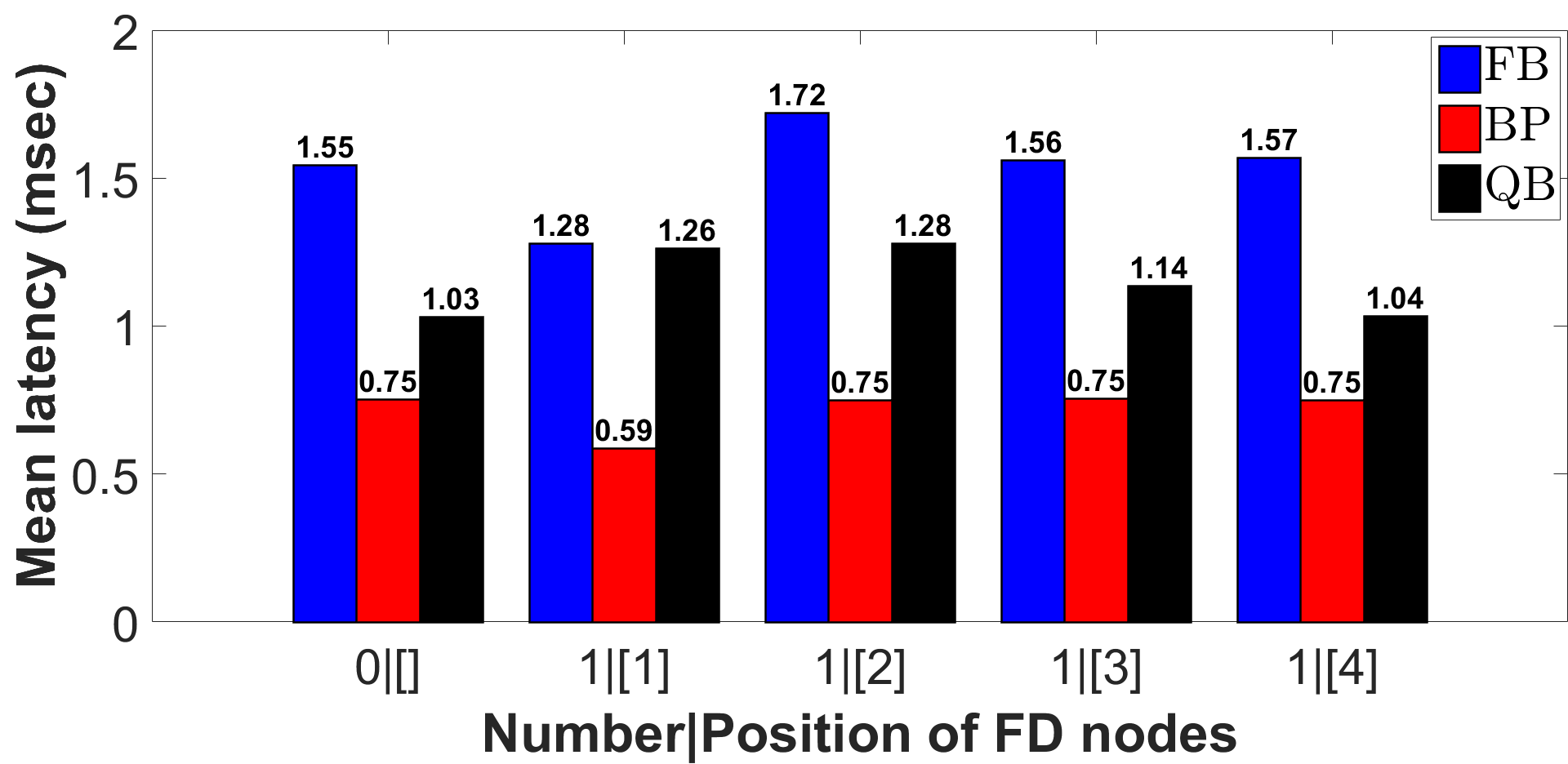}
	\caption{Mean latency for input rate of 0.04 packets/slot and a single FD vehicle - SIC = 40 dB}
	\label{fig:FDmov}
\end{figure}

In Fig. \ref{fig:FDmov} there is a single FD vehicle in the platoon in different positions. In this figure, the importance of the position of the FD-enabled vehicles is illustrated, which depends on the traffic pattern. Particularly, the best option is to have a single FD vehicle in position 1, since this vehicle relays packets from all flows. One FD vehicle in position 3 or 2 cannot provide significant advantage, as the vehicle in position 1 is HD and, hence, it will always "slow down" its receiving packets. Finally, when there is a single FD vehicle, QB scheme always outperforms the FB scheme and it also keeps a small performance gap from BP for most FD nodes positions.
\section{Conclusions}\label{SecV}
In this paper, we have proposed extensions to three scheduling algorithms to cover the case of having FD vehicles in a platoon network where the latency performance was presented. It was demonstrated that the maximum latency performance of a TDMA scheme such as QB, that takes the information of the queue backlogs into account, delivers almost the same results with BP in the HD case, but with significantly reduced signalling overhead and complexity. We have verified that the practical SIC level of 40 dB already allows low packet delay. As a future step, distributed scheduling algorithms will be considered for more complex vehicular networks.

\bibliographystyle{IEEEtran}
\bibliography{IEEEabrv, library}

\end{document}